\documentclass[doublecol]{epl2}
\usepackage{graphicx}
\newcommand{\beq}{\begin{equation}}
\newcommand{\eneq}{\end{equation}}

\begin{document}

\title{Pairing of Cooper pairs in a Josephson junction network containing
an impurity}

\author{D. Giuliano \inst{1} \and P. Sodano \inst{2} }

\institute{\inst{1} Dipartimento di Fisica, Universit\`a della Calabria
             and I.N.F.N., Gruppo Collegato di  Cosenza, Arcavacata di
             Rende, I-87036, Cosenza, Italy
              \\
  \inst{2} Dipartimento di Fisica, Universit\`a di Perugia
             and I.N.F.N., Sezione di Perugia, Via A. Pascoli, I-06123, Perugia,
             Italy}


\abstract{We show how to induce pairing of Cooper pairs (and, thus, $4e$
superconductivity) as a result of local embedding of a quantum
impurity in a  Josephson network fabricable with conventional
junctions. We find that a boundary double Sine-Gordon model
provides an accurate description of the dc Josephson current
patterns, as well as of the stable phases accessible to the
network. We point out that tunneling of pairs of Cooper pairs is
robust against quantum fluctuations, as a consequence of the time
reversal invariance, arising when the central region of the
network is pierced by a dimensionless magnetic flux
 $\varphi = \pi$.  We find that, for $\varphi = \pi$, a stable
attractive finite coupling fixed point emerges and point out its
relevance for engineering
 a two level quantum system with enhanced coherence.}

\maketitle

There is a large number of physical systems that can be mapped
onto quantum impurity models in one dimension  \cite{boundary}.
Embedding a quantum impurity in a condensed matter system may
alter its responses to external perturbations \cite{hewson},
and/or induce the emergence of non Fermi liquid, strongly
correlated phases \cite{nonfl}. In quantum devices with tunable
parameters impurities may be realized by means of point contacts,
of constrictions, or by the crossing of quantum wires or Josephson
junction chains \cite{beenakker1,beenakker2,aoc,giuso4}. For
instance, novel quantum behaviors have been recently evidenced in
the  analysis of $Y$-junctions of quantum wires \cite{aoc} and of
Josephson junction (JJ) chains \cite{giuso4}. Here, we show how
embedding a pertinent impurity in a JJ-chain may lead to the
emergence of nontrivial symmetry protected quantum phases
associated \cite{doucout1, iofe, cataudella,conjun} with the emergence of
4e superconductivity in the network.

While a standard perturbative approach works fine when  impurities
are weakly coupled to the other modes of the system (the
``environment''), there are situations in which  the impurities
are strongly coupled to the environment, affecting its behavior
through a change of  boundary conditions: when this happens,  it
is impossible to disentangle the impurity from the rest of the
system, the perturbative approach breaks down, and, consequently,
one has to resort to nonperturbative methods, to study the system
and the impurity as a whole. Such nonperturbative tools are
naturally provided by boundary  field theories (BFT)
\cite{boundary}: BFTs  allow for deriving exact, nonperturbative
informations from simple, prototypical models which, in many
instances, provide an accurate description of experiments on
realistic low dimensional systems \cite{giamarchi}. In particular,
BFTs have been succesfully used to describe the dc Josephson
current pattern in Josephson devices, such as chains with a weak
link \cite{glark,giuso1}, and SQUIDs \cite{glhek,giuso2}.

\begin{figure}
\includegraphics*[width=1.0\linewidth]{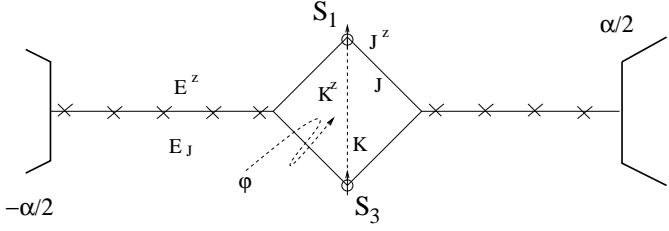}
\caption{The network.} \label{rhombus}
\end{figure}

In this letter, we analyze a Josephson junction network (JJN),
whose BFT description is given by a boundary double Sine-Gordon
model (DSGM) \cite{doublesg}. The device is made by two half JJ
chains, joined through a weak link to a central Aharonov-Bohm cage
{\bf C} \cite{cage}, pierced by a (dimensionless) flux $\varphi =
\Phi / \Phi_0^* (\Phi_0^* = h c / ( 2e))$ (see Fig.\ref{rhombus}).
For simplicity, we connect the outer ends of the chains to two
bulk superconductors at fixed phase difference $\alpha$.

\begin{figure}
\includegraphics*[width=1.0\linewidth]{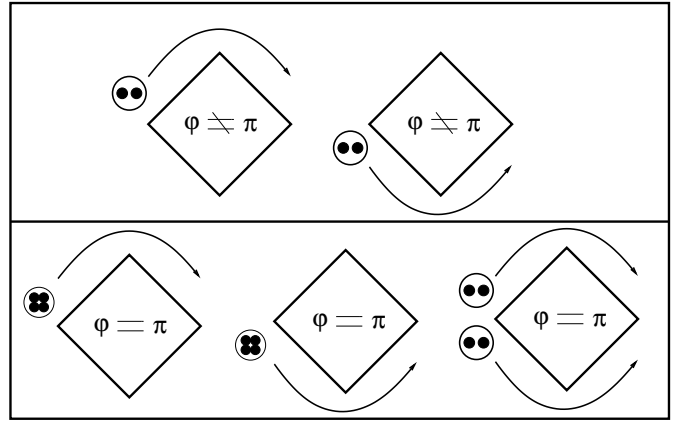}
\caption{Leading tunneling processes across {\bf C}:
 {\bf Top panel}: Single
Cooper pair tunneling for $\varphi \neq \pi$;
{\bf Bottom panel}: Tunneling of pairs of Cooper pairs for
$\varphi = \pi$.}
\label{topa}
\end{figure}

As we shall see, the dc Josephson current across the JJN is a
periodic function of $\alpha$ of period $2 \pi$, given by $I_J (
\alpha ) = \bar{E}_W^{(2)} ( \varphi ) \sin ( \alpha ) + 2
\bar{E}_W^{(4)} ( \varphi ) \sin ( 2 \alpha )$. Varying $\varphi$
changes the ratio $ \bar{E}_W^{(4)} ( \varphi ) / \bar{E}_W^{(2)}
( \varphi )$; in particular, for $\varphi = \pi$, $\bar{E}_W^{(2)}
( \pi )  = 0$, and, thus, the period of $I_J ( \alpha )$ is
halved, signalling the tunneling of pairs of Cooper pairs (PCP)s
through the JJN. A semiclassical analysis \cite{propov} already
accounts for the two harmonics contributing to  $I_J ( \alpha )$:
though, for a generic $\varphi$, single Cooper pair (SCP)
tunneling is the dominating process for charge transport across
the JJN (see top panel of Fig.\ref{topa}), for $\varphi = \pi$,
disruptive interference across  {\bf C} forbids SCP tunneling, and
only allows for   PCP tunneling, thus letting $4e$
superconductivity to emerge in the JJN (see bottom panel of
Fig.\ref{topa}).

It is well known \cite{glhek} that semiclassical and/or mean field
approaches break down in one-dimensional JJNs, when the quantum
phase fluctuations diverge logarithmically with the length of the
system thus inducing a nonperturbative renormalization of the
Josephson couplings. Here, we provide a full quantum treatment of
the JJN in Fig.\ref{rhombus}, based on the BFT approach: We shall
derive the dc Josephson current patterns and evidence the
emergence-  for a suitable choice of the control and constructive
parameters of the JJN- of a robust $4e$ superconductivity
associated to a new attractive fixed point in the phase diagram
accessible to the network.

The central region is described by
\beq H_{\bf C} = \frac{E_c}{2}
\sum_{ j = 0}^3 [ {\bf Q}_j ]^2
 - 2 J \sum_{ j = 0}^3 \cos ( \chi_{j,j+1})
+ J^z \sum_{ j =0}^3{\bf Q}_j {\bf Q}_{j+1} \:\:\:\: ,
\label{rho1} \eneq \noindent with ${\bf Q}_j =  \left[ - i\frac{
\partial }{
\partial \chi_j} -  V_g \right]$, and $ \chi_{j,j+1} =  \chi_{j}- \chi_{j+1}
+ \frac{\varphi}{4}$. $\chi_j$ is the phase of the superconducting
order parameter at the island $j$, $V_g$ is a gate voltage applied
to each superconducting island, $J$ is the Josephson energy of
each junction,
 $E_c$ is the charging energy at each site, and $J^z$ accounts for Coulomb
repulsion  between charges on nearest neighboring junctions. In
the charging regime, that is,  $\frac{J}{E_c} , \frac{J^z}{E_c}
\ll 1$, and tuning $V_g$ so that  $V_g = {\cal N} + \frac{1}{2}$,
with integer ${\cal N}$, one may write $H_{\bf C}$ as a spin-1/2
XXZ-model \cite{giuso1}, whose Hamiltonian is given by $H_{\bf C}
= - J  \sum_{ j = 0}^3 \{ e^{ i \frac{\varphi}{4} } S_j^+
S_{j+1}^- + {\rm h.c.} \} + J^z  \sum_{ j = 0}^3  S_j^z
S_{j+1}^z$. Using the standard bosonization approach
\cite{giamarchi}, the  JJN Hamiltonian may be described by two
spinless Luttinger liquids (LL), interacting with isolated
spin-1/2 variables on {\bf C}, with Hamiltonian given by \beq
H_{\rm JJN} = H_{LL} [ \{ \Phi_a \}  ] + H_T [ \{ \Phi_a\}  , \{
\Theta_a \}; {\bf S}_1 , {\bf S}_3 ] + H_K [ {\bf S}_1 , {\bf S}_3
] \:\:\:\: , \label{rho5} \eneq \noindent with the Luttinger
liquid Hamiltonian $H_{LL} [ \{ \Phi_a \}  ] = \frac{g}{ 4 \pi}
\int_0^L  d x  \sum_{ a = L , R} \left[ \frac{1}{u} \left( \frac{
\partial \Phi_a}{ \partial t} \right)^2 + u \left( \frac{ \partial
\Phi}{ \partial x} \right)^2 \right]$, the interaction Hamiltonian
$H_T [ \{ \Phi_a \} , \{ \Theta_a \} ; {\bf S}_1 , {\bf S}_3 ]= -J
\sum_{a = L,R; s =1,3}  [ e^{ \frac{i }{\sqrt{2}} \Phi_a ( 0 ) }
 e^{ i \epsilon_a \frac{ \varphi}{4}} S_s^- + {\rm h.c.}]$  +
$\sum_{a = L,R; s =1,3}
 \frac{ J_z }{ \sqrt{2} \pi}
 \frac{ \partial \Theta_a ( 0 )}{ \partial x} S_s^z$ ( $\epsilon_L = 1 ,
\epsilon_R = -1$), and $ H_K [ {\bf S}_1 , {\bf S}_3 ] - K [ S_1^+
S_3^- + S_1^+ S_3^- ] + K_z S_1^z S_3^z$. ${\bf S}_1,{\bf S}_3$ are
defined as in Fig.\ref{rhombus}, $\Phi_a$ is the
collective plasmon field of the half chain $_a$, while  $\Theta_a
( x , t )$ is its dual field, that is, $\frac{
\partial \Phi_a ( x , t )}{ \partial x} = \frac{1}{g} \frac{ \partial \Theta_a
( x , t )}{ u \partial t}$, and $ \frac{ \partial \Phi_a ( x , t
)}{ u \partial  t } = \frac{1}{g} \frac{ \partial \Theta_a ( x , t )}{
\partial x}$. The LL parameters are defined as $ g = \pi  / [ 2 (
\pi - {\rm arccos} ( \frac{\Delta}{2} ) ) ]$, $u  = v_f
\frac{\pi}{2} ( \sqrt{1 - (\frac{\Delta}{2})^2} ) / ( {\rm arccos
( \frac{\Delta}{2} )} )$, with $\Delta = E^z / E_J$, $v_f = 2 a
E_J$, where $E_J$ and $E^z$ are the Josephson energy and the
Coulomb repulsion energy of the half chains, and  $a$ is the
lattice step \cite{shulz}. In deriving Eq.(\ref{rho5}), it is
assumed that $J / E_J \ll 1$ and  $J^z / E^z \ll 1$ (i.e., that
{\bf C} is weakly coupled to the chains); this allows to use
Neumann boundary conditions at $x = 0$, that is $\frac{ \partial
\Phi_L ( 0 ) }{ \partial x} = \frac{ \partial \Phi_R  ( 0 )}{
\partial x} = 0$.

The couplings  $K , K_z$ in $H_T  [ \Phi_L , \Phi_R , \Theta_L ,
\Theta_R ; {\bf S}_1 , {\bf S}_3 $ are dynamically generated by
the interaction between {\bf C} and the half chains, as it may be
easily inferred from  the  renormalization group (RG) equations
for the dimensionless  couplings ${\cal K} = L K , {\cal K}^z = L
K^z$, and   ${\cal J} = L^{ 1 - \frac{1}{2g}} J ,  {\cal J}^z =
J^z$. Indeed, employing standard BFT techniques
\cite{cardy,giuso3}, one obtains $\frac{ d {\cal K} }{ d \ln (
\frac{L}{L_0} )}  = {\cal K} + \left[ 1 + \cos ( \varphi ) \right]
( {\cal J} )^2$ , $\frac{ d {\cal K}^z }{ d \ln ( \frac{L}{L_0} )}
= {\cal K}^z + ( {\cal J}^z )^2$, and $\frac{ d {\cal J} }{ d \ln
( \frac{L}{L_0} )}  = \left[ 1 - \frac{1}{2g} \right] {\cal J},
\frac{ d {\cal J}^z }{ d \ln ( \frac{L}{L_0} )} \approx 0$,
showing that both ${\cal K}$ and ${\cal K}^z$ are dynamically
generated whenever $\varphi \neq \pi$.

Integrating  the RG equations one sees  that ${\cal J} (L) \sim
L^{  1 - \frac{1}{2g}}$,  while ${\cal K} ( L) \sim L^{ 2 -
\frac{1}{g}}$, that is, for  $\varphi \neq \pi$, ${\cal K}$ is
always more relevant than ${\cal J}$. At variance, for   $\varphi
= \pi$, no $K$-coupling is generated and  ${\cal J}$ is the only
relevant coupling strength. For $g >  1 / 2$, the BFT description
of the JJN allows  to make very general statements regarding the
regimes accessible to a network of finite size L. Namely, there
will be a perturbative weak coupling regime, accessible for small
${\cal K}$ and ${\cal J}$, and a non-perturbative strong coupling
regime, accessible when ${\cal K}$ or ${\cal J}$ becomes $\sim 1$.
Most importantly, there will be a renormalization group invariant
length scale $L_* \sim  J^{- \frac{1}{2g-1}}$,
 such that for $L < L_*$, the JJN  is in the perturbative weak
coupling regime, while it is in the nonperturbative strong
coupling regime for $L > L_*$.

To account for the last term of Eq.(\ref{rho5}), one may resort to
a Schrieffer-Wolff transformation \cite{swolff}, which amounts to
projecting over the ground state  of {\bf C}, by summing over its
high energy states. At $\varphi = \pi$, the ground state of {\bf
C} is twofold degenerate, while such a degeneracy disappears at
$\varphi \neq \pi$. As a result, for $\varphi \neq \pi$, the
central region {\bf C} is effectively described by the boundary
Hamiltonian $H_{\bf B}$, given by \beq H_{\bf B} [ \varphi ]
 = -  E_W^{(2)} ( \varphi ) \cos [ \Phi_- ( 0 ) ] -
  E_W^{(4)} ( \varphi ) \cos [ 2 \Phi_- ( 0 ) ]
\:\:\:\: , \label{rkky9bis} \eneq \noindent with $E_W^{(2)} (
\varphi) = \cos ( \frac{\varphi}{2} ) \frac{2 J^2}{ K + K^z} +
\sin^2 ( \frac{ \varphi}{2} ) \cos ( \frac{\varphi}{2} ) \frac{ 2
J^4}{  K ( K + K^z )^2}$,
 $E_W^{(4)} ( \varphi)
 = \sin^2 ( \frac{ \varphi}{2} ) \frac{ 2 J^4}{  K ( K + K^z )^2}$,
while $\Phi_- = [ \Phi_L - \Phi_R ] / \sqrt{2}$. At variance, for
$\varphi = \pi$, $H_{\bf B} [ \pi ] \sim
 - \frac{ (E_J)^4}{  J^3} \cos [ 2 \Phi_- ( 0 ) ]$

\begin{figure}
\includegraphics*[width=1.0\linewidth]{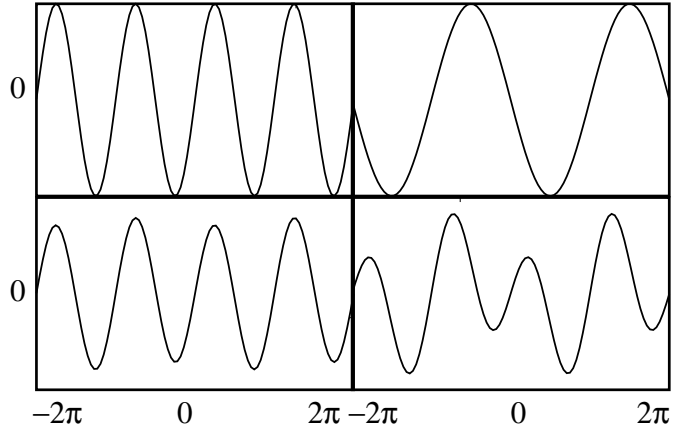}
\caption{Josephson current {\it vs.} $\alpha$ for different values
of $\varphi$ and for $ J^2 / K^2 \approx 0.3$: {\bf Top left
panel} $\varphi = \pi$,  {\bf Bottom
 left panel} $\varphi = 1.01\pi$,{\bf Bottom
 right panel} $\varphi = 1.1 \pi$, {\bf Top
 right panel} $\varphi = 2 \pi$ }
\label{current}
\end{figure}

$I_J ( \alpha )$ may be perturbatively computed as
 $I_J ( \alpha ) = \lim_{\beta \to \infty} -\frac{1}{\beta}
\frac{\partial \ln {\cal Z} }{ \partial \alpha} $, where ${\cal
Z}$ is the partition function of the JJN, given by $\ln {\cal Z}
\approx \ln {\cal Z}_0 - \int_0^\beta \; d \tau \: \langle H_{\bf
B} ( \tau ) \rangle$,  $H_{\bf B} ( \tau )$ being is the boundary
interaction Hamiltonian in the (imaginary time) interaction
representation, while ${\cal Z}_0$ is the partition function at
$H_{\bf B} = 0$. From Eq.(\ref{rkky9bis}), one obtains \beq I_J (
\alpha ) = \bar{E}^{(2)}_W ( \varphi) \sin ( \alpha ) + 2
\bar{E}^{(4)}_W  ( \varphi)\sin ( 2 \alpha ) \;\;\;\; ,
\label{currentj1} \eneq \noindent with $\bar{E}^{(2)}_W ( \varphi)
 = \left( \frac{a}{L} \right)^{\frac{1}{g}} E^{(2)}_W  ( \varphi),
\: \bar{E}^{(4)}_W  ( \varphi)
 = \left( \frac{a}{L} \right)^{\frac{4}{g}} E^{(4)}_W ( \varphi) $.
The ratio $ \bar{E}_W^{(2)} ( \varphi ) / \bar{E}_W^{(4)} (
\varphi )$, measures
 the relative weight of SCP {\it vs.} PCP tunneling
rate. One  notices that, for $\varphi = \pi$, $ \bar{E}_W^{(2)} (
\pi ) = 0$, while $\bar{E}_W^{(4)} ( \pi ) \neq 0$. Thus, at
$\varphi = \pi$ PCP tunneling is  the only allowed mechanism for
charge transfer across {\bf C}. In Fig.(\ref{current}), we plot
$I_J ( \alpha )$ {\it vs.} $\alpha$ for different values of
$\varphi$. For $\varphi \neq \pi$, $I_J ( \alpha )$ is periodic,
with period $\Delta \alpha = 2 \pi$, while, for $\varphi \sim
\pi$, the period shrinks to $\Delta \alpha = \pi$.

For   $g>1$, $H_{\bf B}$ is a relevant operator. Thus, when $L/L_*
\geq 1$, the boundary Hamiltonian  in Eq.(\ref{rkky9bis}) is the
dominating potential term, and the field $\Phi_- (0, \tau)$ takes
values corresponding to minima of $H_{\bf B}$; it obeys Dirichlet
boundary conditions at both boundaries, yielding the mode
expansion $\Phi_- ( x , \tau ) = \alpha + \left( \frac{L - x}{L}
\right) \pi P + \sqrt{\frac{2}{g}} \sum_{ n \neq 0} \sin \left(
\frac{\pi n x}{L} \right) \frac{\alpha_n}{n} e^{ - \frac{\pi n u
\tau}{L}}$  \cite{giuso1}. At the strongly interacting fixed
point, instanton trajectories are the leading quantum
fluctuations; they are described by  imaginary time trajectories $
P ( \tau )$, with $  P ( \tau \to - \infty )$ and  $P ( \tau \to
\infty )$ corresponding to  nearest neighboring  minima
 of $H_{\bf B}$.

\begin{figure}
\includegraphics*[width=1.0\linewidth]{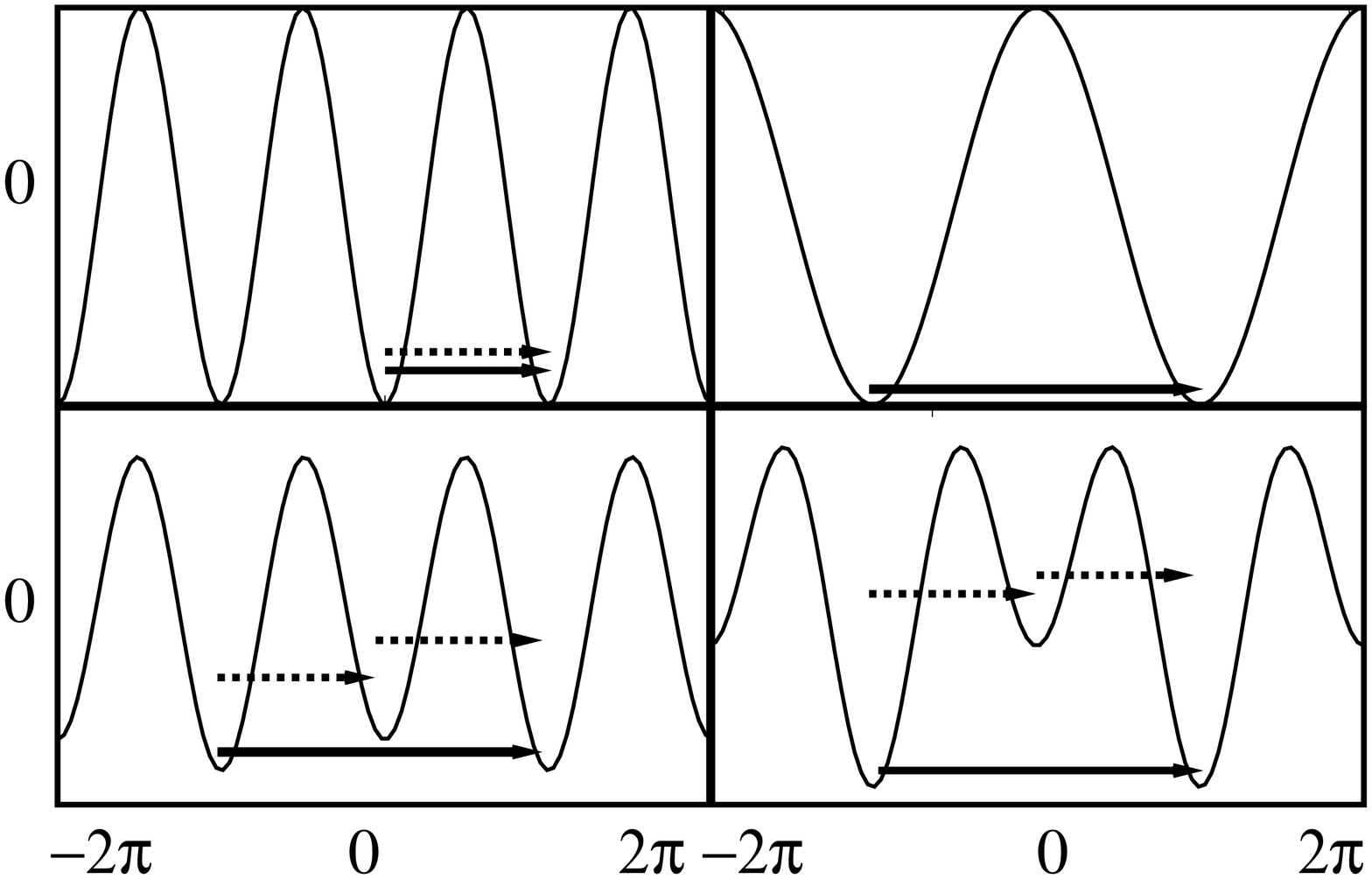}
\caption{Minima of the boundary potential ${\cal H}_{\bf B}$ for
different values of $\varphi$ and for $\Lambda^{(2)} ( \pi ) /
\Lambda^{(4)} ( 0 )  \approx 0.3$ (in arbitrary units). The long
(short) instanton trajectories are represented  as solid(dashed)
arrows: {\bf Top left panel} $\varphi = \pi$,  {\bf Bottom
 left panel} $\varphi = 1.01 \pi$,{\bf Bottom
 right panel} $\varphi = 1.1 \pi$, {\bf Top
 right panel} $\varphi = 2 \pi$ }
\label{minima}
\end{figure}

The instanton profile is derived from $\frac{ \delta S_{\rm Eff} [
P ] }{ \delta P ( \tau )} = 0 $, where  $S_{\rm Eff} [ P ]= \int [
\prod_n {\cal D}  \alpha_n ] \: e^{ - S_E [ \Phi_- ] }$ is
obtained from the Euclidean   action $S_E [ \Phi_- ]$ for a
spinless LL with parameters $g$ and $u$, with a boundary
interaction Hamiltonian given by ${H}_{\bf B}$, after integrating
over the oscilator
 modes of $\Phi_- ( x , \tau )$, $\{ \alpha_n \}$.
 As a result, one gets
\beq
 S_{\rm Eff} [ P ] = \int_0^\beta \: d \tau \: \biggl\{ \frac{M}{2} ( \dot{P} )^2
+ \frac{g \pi u}{4 L } P^2 + V [ \pi P] \biggr\} \:\:\:\: ,
\label{insta5} \eneq \noindent where the ``instanton mass'' $M
\sim \ln ( u {\cal T} / a)$, ${\cal T}$ being the ``instanton
size'', while $V ( x ) =  -  E_W^{(2)} ( \varphi )
 \cos (x) -  E_W^{(4)} ( \varphi) \cos ( 2 x )$. From the effective
action in Eq.(\ref{insta5}), one gets the equation of motion $M
\ddot{P} -  \frac{g \pi u}{2 L }  P +  \pi  E_W^{(2)} ( \varphi )
\sin [ \pi P ] +2 \pi E_W^{(4)} ( \varphi) \sin [ 2 \pi P ] = 0$
which, neglecting the  ``inductive term'' $\propto \frac{1}{L}$,
describes soliton solutions of  the double-Sine Gordon model
\cite{doublesg}. These are given by $P ( \tau ) = \sum_{a=\pm 1}
\frac{2}{\pi} \tan^{-1} \left[ \exp \left( \frac{2 \pi}{ \sqrt{M}}
( a \tau + R ( \varphi )  ) \right) \right]$, with $R ( \varphi )
$  defined by $\frac{1}{4} \sinh^2 \left[ \left( | E_W^{(2)} ( \varphi ) | +
\frac{E_W^{(4)} ( \varphi) }{2} \right) R ( \varphi ) \right] = 2E_W^{(4)} (
\varphi ) / | E_W^{(2)} ( \varphi ) | $.

\begin{figure}
\includegraphics*[width=1.0\linewidth]{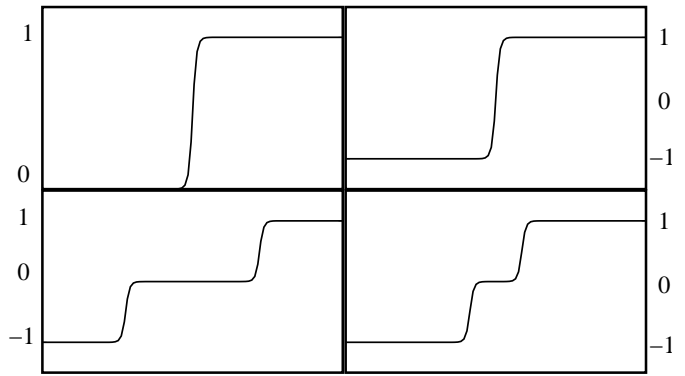}
\caption{Profile of the instanton excitations $P ( \tau )$ for
different values of $\varphi$ and for $J^2 / K^2 \approx 0.3$,
$M=1$: {\bf Top left panel} $\varphi = \pi$,  {\bf Bottom
 left panel} $\varphi = 1.01 \pi$,{\bf Bottom
 right panel} $\varphi = 1.1 \pi$, {\bf Top
 right panel} $\varphi = 2 \pi$ }
\label{instanton}
\end{figure}

In Fig.\ref{minima}, we plot ${H}_{\bf B} [ \Phi] $ {\it vs.}
$\Phi$ for various values of $\varphi \in [ \pi , 2 \pi ]$ and for
$- 2 \pi \leq \Phi ( 0 ) \leq 2 \pi$. We see that, for $\varphi
\neq \pi$, the minima are separated by a distance $2 \pi$. The
corresponding instanton trajectories correspond to a ``long jump''
from $P ( - \infty ) = -1$  to $P ( \infty )=1$ (represented by a
solid arrow in Fig.\ref{minima}). This may be regarded as a
sequence of two ``short instantons'' (dashed arrows), separated by
a distance  $ 2R ( \varphi )$ (see Fig.\ref{instanton}). As
$E^{(2)}_W ( \varphi )$ becomes smaller (that is, as $\varphi$
gets closer to $\pi$), $ R ( \varphi )$ increases,
 and eventually diverges, as $\varphi \to \pi^+$.

Though the degeneracy between the minima is broken by the
$1/L$-term, a pertinent tuning of the phase difference $\alpha$
allows to restore it. Indeed, for $\varphi
> \pi$, the degeneracy of the minima at $\Phi_- ( 0 ) = 2 \pi n
\pm \pi$ is restored by choosing $\alpha = 2 \pi n , n \in Z$,
while, for $\varphi < \pi$, one has to choose $\alpha = \pi(1+2
n)$ and, at $\varphi = \pi$, $\alpha =  \pi ( \frac{1}{2} +2 n)$.

In a BFT approach, long instantons  are represented by the
boundary vertex operators $e^{ \pm i \Theta ( 0 , \tau)}$, with
scaling dimension $h_1 =g$, while short instantons by  the
operators $e^{ \pm \frac{i}{2} \Theta ( 0 , \tau)}$, with scaling
dimension $h_{\frac{1}{2}} = \frac{g}{4}$. For $1<g<4$ and
$\varphi = \pi$, short instantons are relevant perturbations of
the strong coupling (Dirichlet) fixed point. As it happens for
$Y$-shaped JJNs, also here a new, time-reversal invariant,
attractive finite coupling fixed point emerges in the boundary
phase diagram \cite{giuso4}, as a result of the twofold degeneracy
of the ground state of {\bf C}. At variance, for $1<g$ and
$\varphi \neq \pi$, no finite coupling fixed point emerges, since,
now, a possible departure from the Dirichlet fixed point should be
due to long instantons, which are an irrelevant perturbation. For
$\varphi = \pi$, the twofold degeneracy of the ground state of
{\bf C} is due to a ${\bf Z}_2$ symmetry of $V [ \Phi_- ] $
manifesting its invariance under time reversal $\varphi
\longrightarrow 2 \pi - \varphi $, while for $\varphi \neq \pi$,
time reversal is not anymore a symmetry of $V [ \Phi_- ] $, as
evidenced in Figs.\ref{minima},\ref{instanton}. Indeed, for
$\varphi = \pi + \delta$, two nearest neighboring minima at the
top left panel of Fig.\ref{minima} are splitted by an amount
$\propto \sin ( \frac{\delta}{2} )$. As a result, for $1<g<4$ and
for $\varphi = \pi$, the  ${\bf Z}_2$ symmetry protects the
robustness of PCP tunneling across {\bf C}.

We used quantum impurities as a resource for inducing {\it local}
$4e$ superconductivity in a Josephson network, fabricable with
conventional Josephson junctions. The proposed tunneling
mechanism, realized for $1<g<4$ and for $\varphi = \pi$, allows
for the emergence of $4e$ superconductivity, as a result of
embedding an impurity in a Josephson network. Our
analysis evidences that the emergence of $4e$ superconductivity is
the signature of the presence in the dc Josephson current across
the device of two distinct- and competing- periodicities, whose
relative weight is tuned by the magnetic flux $\varphi$ piercing
the central region. Since, for a generic value of $\varphi$, the
double sine-Gordon boundary potential has been normalized so that
the Cooper pair charge $2e$ is associated to the higher
periodicity, one gets that, when - at $\varphi$ equal to $\pi$-
the period is halved, the charge carriers should have charge $4e$.
Thus, the phenomenon analyzed in this paper is basically different
from the mere phase difference renormalization taking place, for
instance, in a series array of $N$ equal Josephson junctions
where, although the Josephson current is $\propto \sin ( \alpha /
N ) $, only a single harmonics is present.

Associated with $4e$ superconductivity there is, for $1<g<4$, a new finite coupling
attractive fixed point, which allows for the possibility of using
the JJN as a two level quantum system with enhanced coherence
\cite{giuso4,frdec}. Furthermore, PCP tunneling is robust, as a
consequence of the time reversal invariance, realized only for
 $\varphi = \pi$.  The proposed mechanism exhibits intriguing
similarities with the
electron bunching phenomenon \cite{sela}, observed  in shot noise
measurements \cite{heib2} on quantum dots in the Kondo regime.


\begin{thebibliography}{99}


\bibitem{boundary} I. Affleck, Lecture Notes, Les Houches, 2008
                   (arXiv:0809.3474).

\bibitem{hewson} A. C. Hewson, ``The Kondo Problem to Heavy Fermions'',
Cambridge University Press (1997), and references therein.

\bibitem{nonfl} P. Nozier\`es and A. Blandin, J. Phys.{\bf 41},193 (1980);
                D. L. Cox and F. Zawadowski, Adv. Phys. {\bf 47}, 599 (1998).

\bibitem{beenakker1} H. van Houten and C. W. J. Beenakker,  Physics Today
49 (7), 22 (1996).

\bibitem{beenakker2} C. W. J. Beenakker and H. van Houten in
''Nanostructures and Mesoscopic Systems'', W. P. Kirk and M. A.
Reed eds. (Academic, New York, 1992).


\bibitem{aoc} C. Chamon, M. Oshikawa and I. Affleck, Phys. Rev. Lett. {\bf 91},
              206403(2003); M. Oshikawa, C.
  Chamon and I. Affleck, Journal of Statistical Mechanics JSTAT/2006/P02008.

\bibitem{giuso4} D. Giuliano and P. Sodano, New. Jour. of Physics
{\bf 10},093023(2008).

\bibitem{doucout1} B. Doucot and J. Vidal, Phys.Rev.Lett.{\bf 88},227005(2002).

\bibitem{iofe} L.B. Ioffe and M.V. Feigel'man, Phys.Rv.B {\bf 66},
224503 (2002); B. Doucot, M.V. Feigel'man, L.B. Ioffe, Phys. Rev.
Lett. {\bf 90}, 107003 (2003).

\bibitem{cataudella} V. Rizzi, V. Cataudella, and R.
Fazio, Phys.Rev.{\bf  B 73},100502(2006).

\bibitem{conjun} Congjun Wu,
Phys. Rev. Lett.{\bf  95}, 266404 (2005).


\bibitem{giamarchi} T. Giamarchi, ``Quantum Physics in One Dimension",
(Oxford University Press, 2004).


\bibitem{glark} L. I. Glazman and A. I.
Larkin, Phys. Rev. Lett.{\bf 79},3736(1997).

\bibitem{giuso1} D. Giuliano and P. Sodano, Nucl.Phys.{\bf B711},480,(2005).


\bibitem{glhek} F.W.J. Hekking and L.I. Glazman, Phys. Rev. {\bf B 55}, 6551
(1997).

\bibitem{giuso2} D. Giuliano and P. Sodano, Nucl.Phy.{\bf B 770},332(2007).


\bibitem{doublesg} D. K. Campbell, M. Peyrard, and P.
Sodano, Physica D,{\bf 19},165(1986), and references therein.


\bibitem{cage} J. Vidal, R. Mosseri, and B. Doucot, Phys. Rev. Lett.{\bf 81},
               5888 (1998).

\bibitem{propov} I. V. Protopopov and M. V.
Feigel' man, Phys.Rev.{\bf B 70},184519(2004);Phys.Rev.{\bf B
74},064516(2006).



\bibitem{shulz} C. N. Yang and C. P. Yang, Phys. Rev. {\bf 150}, 327 (1966);
For a review see, for instance, H. J. Shulz, G. Cuniberti and P.
Pieri in {\it Field Theories for Low-Dimensional Correlated
Systems}, G. Morandi, P. Sodano, V. Tognetti and A. Tagliacozzo
eds., Springer, Berlin.


\bibitem{cardy} J. Cardy,  Encyclopedia of Mathematical Physics,
                (Elsevier, 2006) (physics arXiv:hep-th/0411189).

\bibitem{giuso3} D. Giuliano and P. Sodano, Nucl.Phys.{\bf B 811},395(2009).


\bibitem{swolff} J. R. Schrieffer and P. A. Wolff, Phys.Rev.{\bf 149},491(1966).


\bibitem{frdec} E. Novais, A. H. Castro Neto, L. Borda, I. Affleck, and
G. Zarand, Phys. Rev.{\bf B 72}, 014417 (2005)

\bibitem{sela} E. Sela, Y. Oreg, F. von Oppen, and J.
               Koch, Phys.Rev.Lett.{\bf 97},86601(2006).

\bibitem{heib2}  O. Zarchin, M. Zaffalon, M. Heiblum, D. Mahalu, and
               V. Umansky, Phys.Rev.{\bf B 77},241303(R)(2008).







\end{thebibliography}
\end{document}